\documentclass[11pt,a4paper]{article}

\usepackage{amsmath}
\usepackage{epsfig}
\usepackage{graphicx}
\usepackage{wrapfig}
\usepackage{epsfig}
\usepackage{psfrag}
\usepackage{substr}
\usepackage[tight]{minitoc}
\usepackage{subfigure}
\usepackage{longtable}
\usepackage{chngcntr}
\usepackage[all]{xy}
\usepackage{cite}
\usepackage{xr}

\begin{document}

\title{Smale--Williams Solenoids in a System of Coupled Bonhoeffer--van der Pol Oscillators}

\author{V.\,M.\,Doroshenko \\
Saratov State Medical University \\
Bolshaya Kazachia 112, Saratov, 410012, Russia \\
\texttt{Dorvalentina9@gmail.com} \\
\and
V.\,P.\,Kruglov \\
Kotel’nikov Institute of Radio Engineering and Electronics, \\
Russian Academy of Sciences, Saratov Branch \\
Zelenaya 38, Saratov, 410019, Russia \\
\texttt{kruglovyacheslav@gmail.com} \\
\and
S.\,P.\,Kuznetsov \\
Kotel’nikov Institute of Radio Engineering and Electronics, \\
Russian Academy of Sciences, Saratov Branch \\
Zelenaya 38, Saratov, 410019, Russia \\
\texttt{spkuz@yandex.ru} \\
}

\maketitle\label{article_begin}
{

\begin{abstract}
The principle of constructing a new class of systems demonstrating hyperbolic chaotic attractors is proposed. 
It is based on using subsystems, the transfer of oscillatory excitation between which is provided resonantly due to the difference 
in the frequencies of small and large (relaxation) oscillations by an integer number of times being accompanied by 
phase transformation according to an expanding circle nap. As an example, we consider a system where a Smale--Williams attractor 
is realized, which is based on two coupled Bonhoeffer--van der Pol oscillators. Due to the applied modulation of parameter controlling 
the Andronov--Hopf bifurcation, the oscillators manifest activity and suppression turn by turn. 
With appropriate selection of the modulation form, relaxation oscillations occur at the end of each activity stage, 
the fundamental frequency of which is by an integer factor $M = 2,3,4, \ldots$ smaller than that of small oscillations. 
When the partner oscillator enters the activity stage, the oscillations start being stimulated by the \textit{M}-th harmonic 
of the relaxation oscillations, so that the transformation of the oscillation phase during the modulation period corresponds to 
the \textit{M}-fold expanding circle map. In the state space of the Poincar\'e map this corresponds to attractor of Smale--Williams type, 
constructed with \textit{M}-fold increase in the number of turns of the winding at each step of the mapping. 
The results of numerical studies confirming the occurrence of the hyperbolic attractors in certain parameter domains are presented, 
including the waveforms of the oscillations, portraits of attractors, diagrams illustrating the phase transformation 
according to the expanding circle map, Lyapunov exponents, and charts of dynamic regimes in parameter planes. 
The hyperbolic nature of the attractors is verified by numerical calculations that confirm absence of tangencies 
of stable and unstable manifolds for trajectories on the attractor (``criterion of angles''). 
An electronic circuit is proposed that implements this principle of obtaining the hyperbolic chaos and its functioning 
is demonstrated using the software package Multisim.
\end{abstract}


\section{Introduction}
\setcounter{equation}{0}
Uniformly hyperbolic attractors represent a special class of attracting sets that consist only of saddle (hyperbolic) trajectories
~\cite{1, 2, 3, 4, 5, 6, 7}. In recurrent maps the tangent space of a saddle trajectory splits into two invariant subspaces. 
One of them is expanding because it consists of vectors with norms exponentially decreasing with backward time evolution. 
Other subspace is contracting because it consists of vectors with norms exponentially decreasing with forward time evolution. 
Rates of decrease are bounded and far from zero at every point of attractor. An arbitrary small perturbation vector of a saddle trajectory 
is a linear combination of vectors belonging to these subspaces. Set of trajectories that asymptotically converge to the reference 
trajectory in forward (backward) time are its stable (unstable) manifold. For the hyperbolic attractors these manifolds can intersect 
only transversely~\cite{1, 2, 3, 4, 5, 6, 7}. 

It was rigorously proved that the uniformly hyperbolic attractors demonstrate chaotic dynamics. The uniformly hyperbolic attractors 
have a number of remarkable features due to their structure~\cite{1, 2, 3, 4, 5, 6, 7}. The most intuitively clear feature is called 
the Sinai--Ruelle--Bowen measure on attractor and manifested in smooth deformation of phase space without formation of local 
discontinuities and singularities. Another important property is that the uniformly hyperbolic attractors, unlike most other chaotic 
attractors, are structurally stable (or rough). The structural stability is qualified by insensitivity of dynamics to small changes 
in the parameters and functions in governing equations. Another feature of the uniformly hyperbolic attractors is a possibility of 
constructing symbolic dynamics corresponding to the dynamics on an attractor. Due to the structure of a uniformly hyperbolic attractor, 
one can establish a one-to-one correspondence between an arbitrary trajectory of an attractor and an infinite sequence of a finite set 
of symbols.

For a long time examples of physical systems with uniformly hyperbolic attractors were not known. Only abstract geometric models, 
such as Smale--Williams solenoid, Plykin attractor and DA--attractor (``derived from Anosov'') were introduced~\cite{1, 2, 3, 4, 5, 6, 7}. 

Smale--Williams solenoids appear as attractors of specially constructed maps in phase space of dimension 3 or more. Let’s say we have 
a toroidal domain in a phase space. One iteration of mapping expands the domain in longitudinal direction with integer factor 
$ \left| M \right| \geq 1 $, 
strongly contracts it in the transverse directions, twists and embeds it inside the initial domain. 

But recently a great amount of examples of physically implementable systems with uniformly hyperbolic attractors were proposed 
~\cite{8, 9, 10, 11, 12, 13, 14}. Mostly they are the Smale--Williams attractors in Poincar\'e cross-sections of certain flow dynamical 
systems. In accordance with the structure of the attractors described above the operation of these systems is based on expanding 
transformation of angular (cyclic) variables, an example of which is the phase of oscillations. This transformation is called Bernoulli 
mapping (or expanding circle map): $ \varphi _{n+1} = M \varphi _{n} + \text{const} \pmod{2 \pi} $. By design there is strong contraction 
in all directions that are transversal to angular.

Historically, the first example of a physically realizable system with Smale--Williams solenoid is a model of two van der Pol oscillators 
with frequencies that differ by factor of $2$ ~\cite{8}. The parameters controlling the Andronov--Hopf bifurcation are periodically 
modulated in such a way that the oscillators get excited alternatively with phases undergoing the Bernoulli mapping with expanding factor 
on each period. This system was studied numerically~\cite{8} and implemented in a radioengineering laboratory device~\cite{9}. The 
hyperbolicity of the attractor was proved in numerical test of expanding and contracting cones~\cite{10}.

We introduce a new principle of construction of systems manifesting hyperbolic attractors and quasiperiodic dynamics. The systems of new 
class consist of oscillators with resonant excitation transfer due to the frequencies of small and large oscillations differing by an 
integer number of times. The phases of oscillations undergo special transformation on full revolution of excitation. We have used 
described approach recently to devise a mechanical system with Smale--Williams attractor. That model was based on two Froude pendulums 
on a common shaft rotating at constant angular velocity with alternating damping by periodic application of the additional friction
~\cite{11}. 

In this article we propose a new model of a system that belongs to the described class. The model consists of two weakly coupled 
Bonhoeffer--van der Pol oscillators with periodically modulated parameters. Oscillators alternatively undergo smooth transition from 
small self-oscillations to relaxation self-oscillations. Depending on parameters, the Smale--Williams solenoids with different expansion 
factors \textit{M} of the angular variable appear in phase space of the system. If one of the parameters is zero, the model becomes a system of 
coupled van der Pol oscillators, in which Smale--Williams solenoids also arise. The absence of an auxiliary signal facilitates the 
implementation of the model as an electronic generator.

We derive a system of ordinary differential equations in Section 1. The results of numerical solution and analysis are in Section 2. We 
suggest an electronic circuit scheme and simulate it in Multisim software in Section 3. 

\section{Definition of the Model}

We start with description of the Bonhoeffer--van der Pol oscillator. The equations of a very similar system were introduced by FitzHugh 
~\cite{15} and the equivalent circuit was proposed by Nagumo et al~\cite{16}. Consider a circuit shown in Fig.~\ref{fig1}. It consists 
of a capacitor \textit{C}, an inductor \textit{L}, a voltage source $ E_0 $, and an element with a non-linear conductivity 
\textit{G}, and differs from the scheme of Nagumo only by neglecting the resistance of one of the branches. The current-voltage 
characteristic of a nonlinear element is supposed to be a third-order polynomial:
 $ g \left( v \right) = -g_1 v \! + \! g_3 v^3 \quad \left( g_1 > 0, \ g_3 > 0 \right) $. 

Differential equations for the voltage on capacitor and for the current through the inductor are obtained with Kirchhoff's rules applied 
to the circuit:

\begin{figure}[!hb]
\centering\includegraphics[scale=0.33,keepaspectratio]{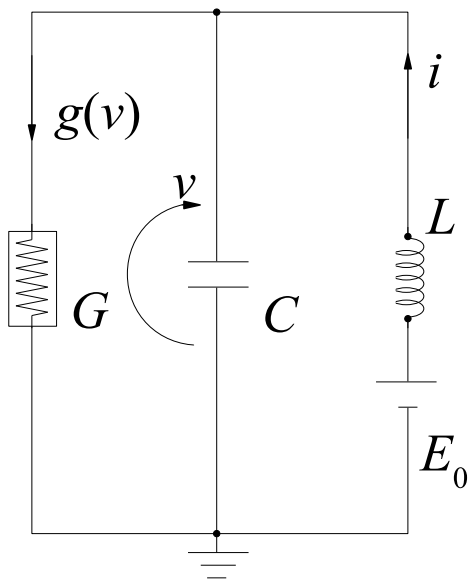} 
\caption{Equivalent scheme of Bonhoeffer--van der Pol oscillator.} 
\label{fig1}
\end{figure}

\begin{equation}
\begin{aligned}
C \frac{dv}{dt} &= i + g_1 v -g_3 v^3, \\ 
L \frac{di}{dt} &= -v + E_0.
\label{eq1}
\end{aligned}
\end{equation}

By the change of variables, the system~\eqref{eq1} reduces to the Bonhoeffer--van der Pol equation, which differs from the van der Pol equation 
by a constant parameter on the right-hand side, where $ x = v \sqrt{3 \omega _0 g_3 C} $ is the new dynamical variable and $ \dot{x} $ is its time derivative, 
$ \omega _0 t \Rightarrow t $ is the new time variable, $ a = \omega _0 g_1 C $ is control parameter, 
$ K = E_0 \sqrt{3 \omega _0 g_3 C} $ is constant and $ \omega _0^2 = \frac{1}{L C}$:

\begin{equation}
\ddot{x} - \left( a - x^2 \right) \dot{x} + x = K,
\label{eq2}
\end{equation}

where $ x = v \sqrt{3 \omega _0 g_3 C} $ is the new dynamical variable and $ \dot{x} $ is its time derivative, 
$ \omega _0 t \Rightarrow t $ is the new time variable, $ a = \omega _0 g_1 C $ is control parameter, 
$ K = E_0 \sqrt{3 \omega _0 g_3 C} $ is constant and $ \omega _0^2 = \frac{1}{L C}$. 

At small positive values of a almost sinusoidal self-oscillations occur with a frequency close to  in the Bonhoeffer--van der Pol 
oscillator. With increase in the parameter a transition to relaxation oscillations takes place, the shape of which differs significantly 
from the sinusoidal, while the basic frequency decreases. If the parameter K is nonzero, then both odd and even harmonics are presented. 
(Note that the spectrum of the van der Pol oscillator contains only odd harmonics.) Fig.~\ref{fig2} presents the dependence of the period 
of self-oscillations on the parameter \textit{a}. The values of the parameter are marked at which almost harmonic self-oscillations with 
the period of $ 2\pi $ and non-harmonic oscillations with the period of $ 4\pi $ are observed. Fig.~\ref{fig3} shows portraits of 
attractors in both regimes. Fig.~\ref{fig4} shows spectra of the oscillations for these regimes. At $ a = 5.539 $, $ K = 0.5 $, the basic 
frequency is approximately $ 1/2 $, so, the frequency of the second harmonic is close to the frequency of small oscillations. This feature 
can be used to construct a system with a Smale--Williams attractor based on two Bonhoeffer--van der Pol oscillators alternately excited by 
a correctly matched periodic modulation of the parameters. 

Fig.~\ref{fig5} shows the resonant excitation of a linear oscillator $ \ddot{y} + y = \varepsilon x $ by the second harmonic of the 
self-oscillating system~\eqref{eq2}. This result is based on a numerical solution with parameters $ a = 5.49 $, $ K = 0.5 $, and 
$ \varepsilon = 0.1 $. Visually clear that the oscillation period of a linear oscillator is twice that of the self-oscillation period 
of the system~\eqref{eq2}, so the frequency of the second harmonic of self-oscillations equals to the natural frequency of the linear 
oscillator. 

\begin{figure}[!ht]
\centering\includegraphics[width=1.\textwidth,keepaspectratio]{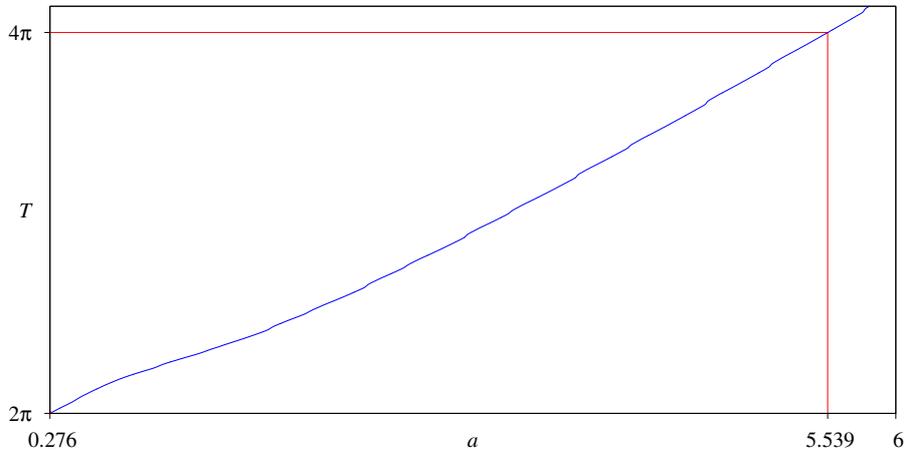} 
\caption{Dependence of the period of self-oscillations of system~\eqref{eq2} on the parameter a at $ K=0.5 $. The values of the parameter 
are marked at which almost harmonic self-oscillations with a period of $ 2\pi $ and non-harmonic oscillations with a period of $ 4\pi $ 
are observed. There are no limit cycles at $ a < K^2 = 0.25 $.} 
\label{fig2}
\end{figure}

\begin{figure}[!h]
\includegraphics[width=.5\textwidth,keepaspectratio]{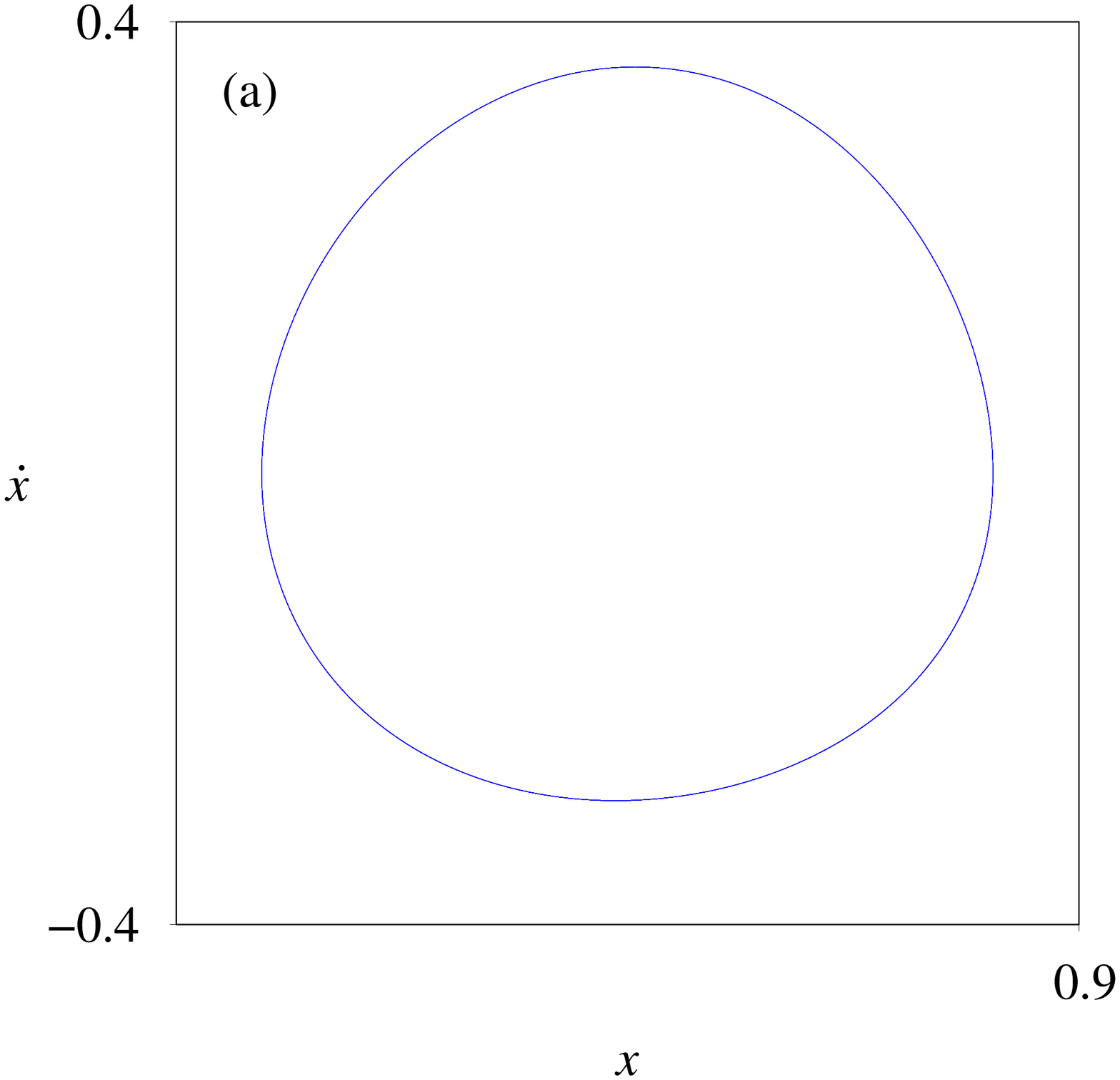}
\includegraphics[width=.5\textwidth,keepaspectratio]{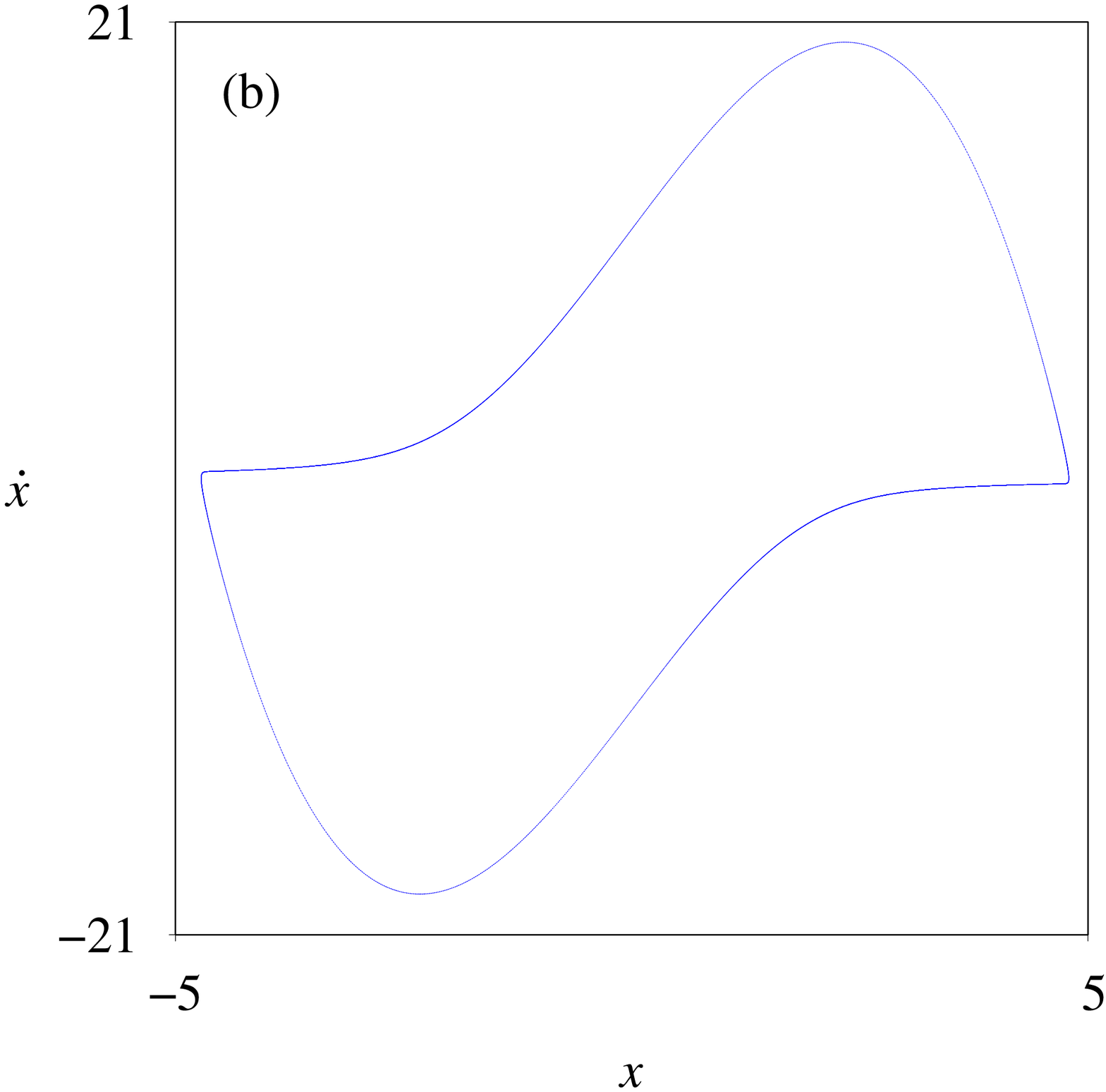}
\caption{Portraits of attractors of the system~\eqref{eq2} at a=0.276 (a) and a=5.539 (b). $ K=0.5 $. }
\label{fig3}
\end{figure}

\begin{figure}[!h]
\includegraphics[width=\textwidth,keepaspectratio]{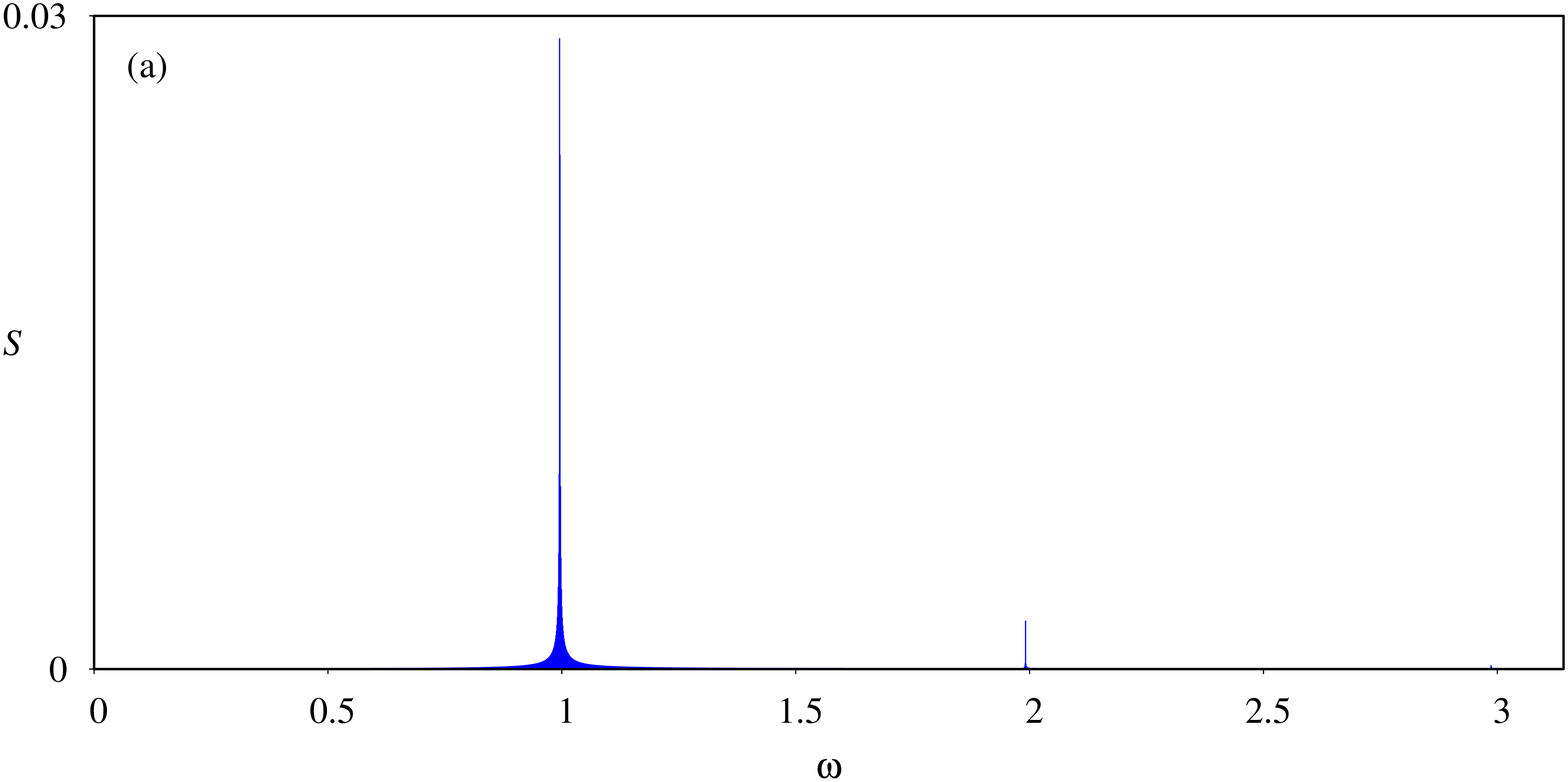} 
\includegraphics[width=\textwidth,keepaspectratio]{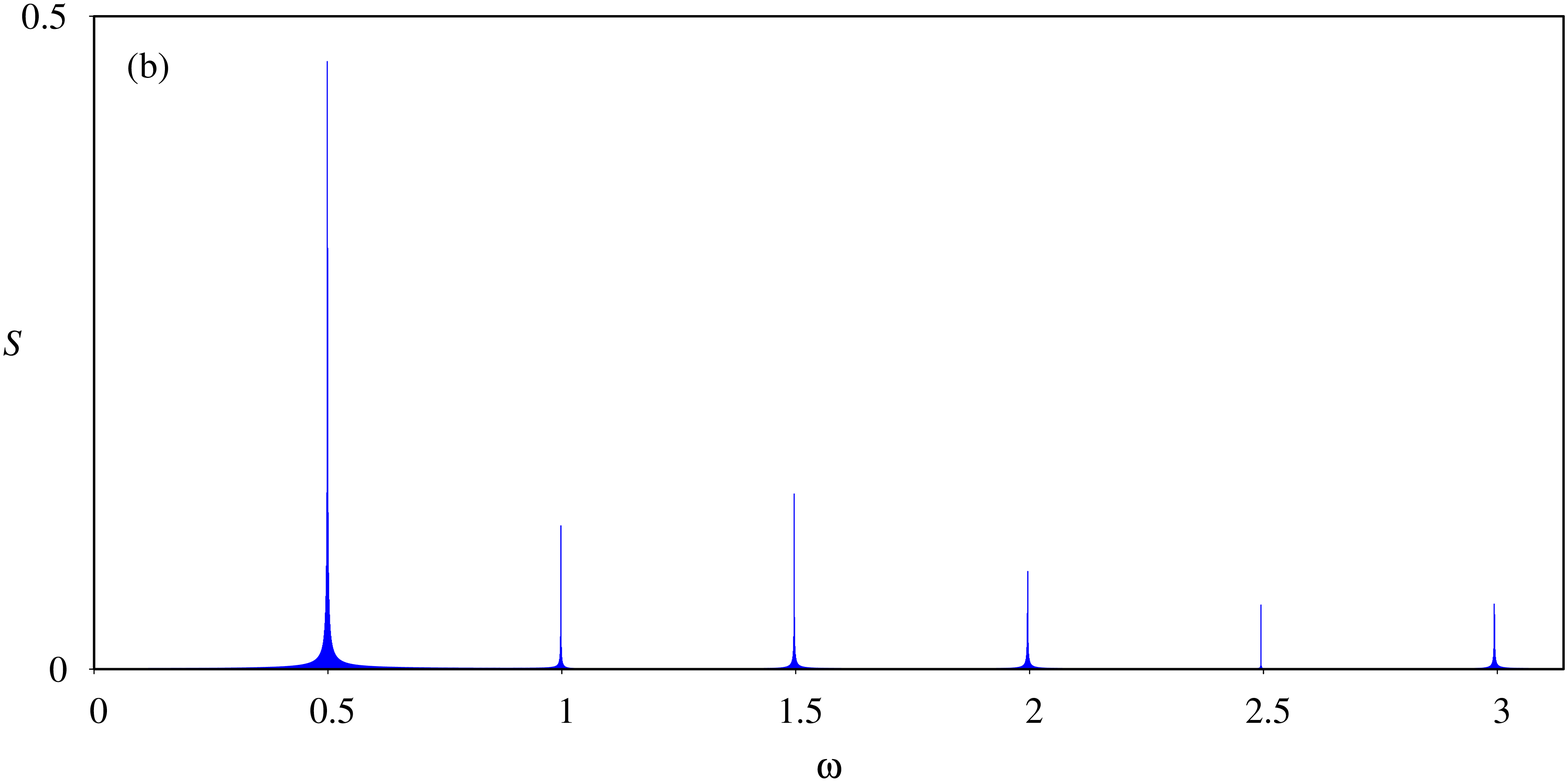} 
\caption{Spectra of oscillations of the system~\eqref{eq2} at a=0.276 (a) and a=5.539 (b). $ K=0.5 $. }
\label{fig4}
\end{figure}

\begin{figure}[!h]
\includegraphics[width=\textwidth,keepaspectratio]{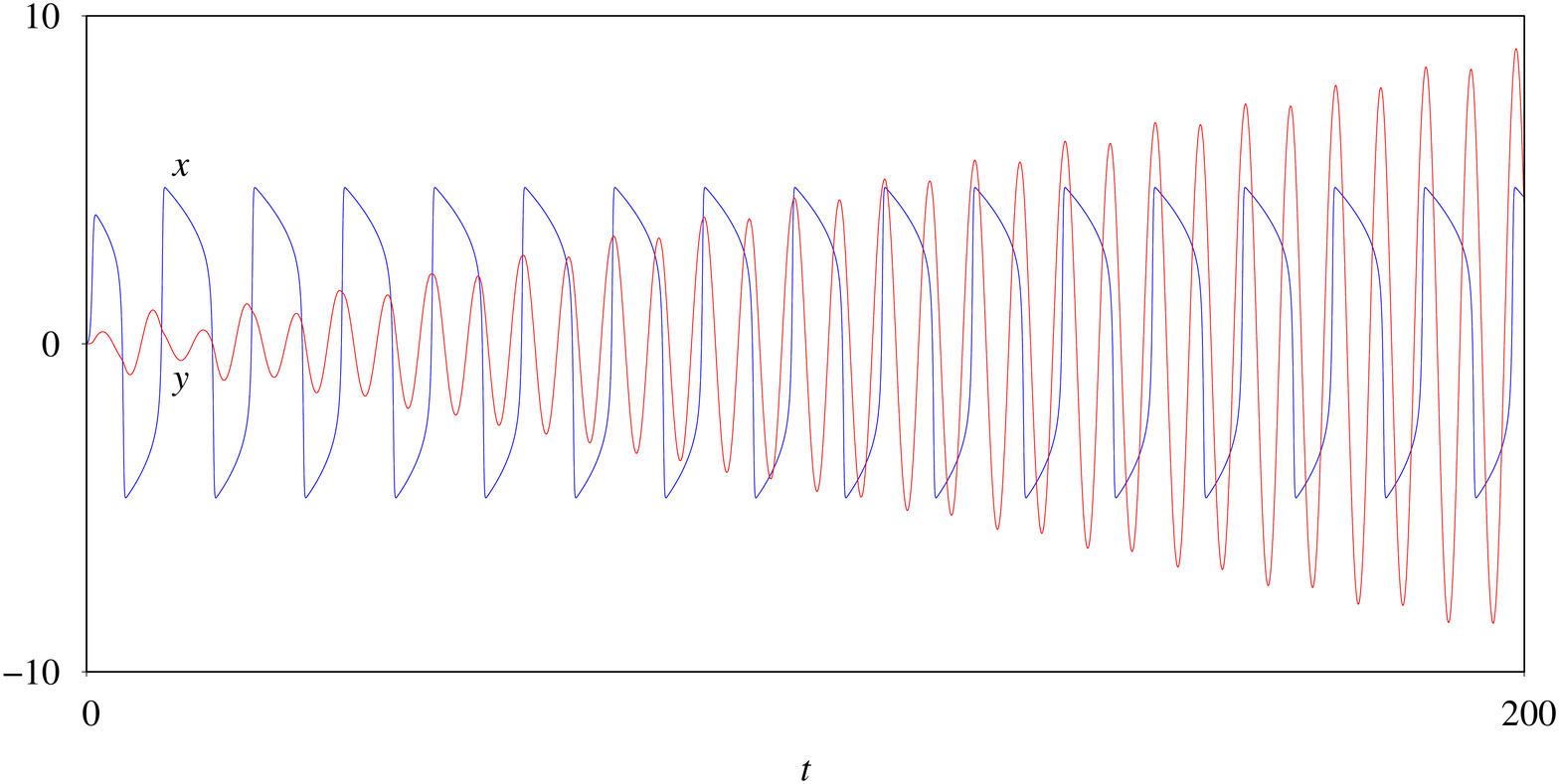} 
\caption{Resonant excitation of a linear oscillator by the second harmonic of the self-oscillating system~\eqref{eq2}. This result is 
based on a numerical solution of the set of equations $ \ddot{x} - \left( a - x^2 \right) \dot{x} + x = K, \ \ddot{y} + y = \varepsilon x $ with parameters $ a=5.49 $, $ K=0.5 $, $ \varepsilon=0.1 $.  }
\label{fig5}
\end{figure}

We propose a system consisting of two identical weakly coupled Bonhoeffer--van der Pol oscillators, in which the control parameters vary 
slowly in time that provides alternating excitation and damping of the oscillations. With increase of parameter \textit{a}, the 
fundamental frequency of the self-oscillations decreases. By modulating the parameters \textit{a} one can obtain a periodic switching of 
the frequency of the subsystems from $ \omega =1 $ for small oscillations to $ \omega = \frac{1}{M} $ (\textit{M} is a natural number, 
$ \left| M \right| > 1 $) for oscillations with a large amplitude. 

Now we describe an appropriate modulation function $ f \left( \tau \right) $. Let the parameter remain constant for a certain time 
$ \left[ 0;\ \tau _1 \right) $ in the excitation stage and equal to the maximum value of \textit{a}, then decrease to a negative value 
($ -c $) on an interval $ \left[ \tau _1 ;\ \tau _2 \right) $, and then increase more slowly on $ \left[ \tau _2 ;\ 1 \right) $, again 
reaching the maximum value of \textit{a}, and suppose the function is of period 1:

\begin{equation}
f \left( \tau \right) = f \left( \tau + 1 \right) = \left\{ \begin{array}{ll}
a, & \textrm{if $ 0 \leq \tau < \tau _1 $,}\\
\frac{ \left( a - c \right) \tau + c \tau _1 - a \tau _2}{\tau _1 - \tau _2} , & \textrm{if $ \tau _1 \leq \tau < \tau _2 $,}\\
\frac{ \left( c - a \right) \tau + a \tau _2 - c}{\tau _2 - 1} , & \textrm{if $ \tau _2 \leq \tau < 1 $.}
\end{array} \right.
\label{eq3}
\end{equation}

With that the model equations are:

\begin{equation}
\begin{aligned}
\ddot{x} - \left( f \left( \frac{t}{T} + \frac{1}{4} \right) -x^2 \right) \dot{x} + x = K + \varepsilon \left( y - x \right), \\ 
\ddot{y} - \left( f \left( \frac{t}{T} - \frac{1}{4} \right) -y^2 \right) \dot{y} + y = K + \varepsilon \left( x - y \right).
\label{eq4}
\end{aligned}
\end{equation}

Modulation functions in both equations are supposed to be shifted on a half-period from each other to provide alternating excitation of 
the oscillators. The oscillators are coupled through the terms $ \varepsilon \left( y - x \right) $ and 
$ \varepsilon \left( x - y \right) $ with small coupling parameter $ \varepsilon $. Parameter \textit{K} is constant. Terms 
$ x^2 \dot{x} $, $ y^2 \dot{y} $ and \textit{K} in the equations ensure the presence of odd and even harmonics of the self-oscillations.

Equivalent alternative form of the equations~\eqref{eq4} is 

\begin{equation}
\begin{aligned}
\dot{x} &= u, \\
\dot{u} &= \left( f \left( \frac{t}{T} + \frac{1}{4} \right) -x^2 \right) u - x + K + \varepsilon \left( y - x \right), \\ 
\dot{y} &= v, \\
\dot{v} &= \left( f \left( \frac{t}{T} - \frac{1}{4} \right) -y^2 \right) v - y + K + \varepsilon \left( x - y \right).
\label{eq5}
\end{aligned}
\end{equation}

Let us discuss operation of the system in regimes with hyperbolic attractors. We start with the situation when one oscillator generates 
self-oscillations, and the second is inhibited. The fundamental self-oscillation frequency, due to the above mentioned selection of 
parameters, may be $ M = 2, \ 3, \ 4, \ \ldots $ times smaller than the frequency of small oscillations. When the second oscillator 
approaches the threshold of excitation and goes over it due to the increase in the control parameter, it excites in a resonant manner 
being stimulated by the action of the spectral component of the first oscillator with frequency $ \omega = 1 $. Therefore, the phase of 
the second oscillator corresponds to the phase of the first oscillator multiplied by \textit{M}. Then the second oscillator approaches 
the steady-state regime of relaxation self-oscillations with a frequency \textit{M} times smaller than the frequency of small 
oscillations and phase \textit{M} times larger then the phase of the first oscillator. Further, the process repeats again and again with 
exchange in the roles of one and the other oscillator. Over a full period of modulation the initial phases of both oscillators are 
multiplied by factor $ M^2 = 4, \ 9, \ 16, \ \ldots $, i.e. the phases undergo the Bernoulli mapping. With strong contraction along other 
directions this corresponds to the dynamics on Smale--Williams solenoids with expanding factor $ M^2 $ generated by Poincar\'e mapping 
$ \textbf{X} _{n+1} = \textbf{F} _T \left( \textbf{X} _n \right) $ over the full period of modulation. 

One can define a mapping $ \textbf{X} _{n+1} = \textbf{F} _{T/2} \left( \textbf{X} _n \right) $ over a half period of modulation 
due to the symmetry $ \left( x, \ y, \ t \right) \leftrightarrow \left( u, \ v, \ t + \frac{T}{2} \right) $ with state vector at the 
instants of time $ t_n = \frac{n T}{2} $ as

\begin{equation}
\textbf{X} _n = \left\{ \begin{array}{ll}
x \left( t_n \right), \ u \left( t_n \right), \ y \left( t_n \right), \ v \left( t_n \right), \  & \textrm{if \textit{n} is odd,}\\
y \left( t_n \right), \ v \left( t_n \right), \ x \left( t_n \right), \ u \left( t_n \right), \  & \textrm{if \textit{n} is even.}
\end{array} \right.
\label{eq6}
\end{equation}

It is easy to implement numerically. In $ \textbf{X} _{n+1} = \textbf{F} _{T/2} \left( \textbf{X} _n \right) $ the phase is 
multiplied by $ M = 2, \ 3, \ 4, \ \ldots $

\section{Results of numerical simulation}

The system of equations~\eqref{eq5} was solved numerically using 4-th order Runge--Kutta algorithm with small enough step. Fig.~\ref{fig6} 
shows the time dependences of the variables \textit{x} and \textit{y}, and the functions governing the modulation of the 
control parameters, with the values $ a = 5.49 $, $ K = 0.5 $, $ c = -2 $, $ \varepsilon = 0.01 $, $ T= 200 $, $ \tau _1 = 0.4 $, 
$ \tau _2 = 0.5 $. One can see that oscillators are excited at some stage of increase of the control parameters, and the characteristic 
period of oscillations also grows with the control parameter. Fig.~\ref{fig7} shows the power density spectrum of the first oscillator 
in logarithmic scale at $ a = 5.49 $, $ K = 0.5 $, $ c = -2 $, $ \varepsilon = 0.01 $, $ T= 200 $, $ \tau _1 = 0.4 $, $ \tau _2 = 0.5 $. 
The spectrum is continuous, which is inherent in chaotic oscillations. The constant component of the signal has the maximum intensity due 
to the constant term \textit{K} on the right-hand side of the equations. The other maxima of the spectrum correspond to frequencies that 
are multiples of $ 1/2 $. An interaction is assumed between components of the spectrum with frequencies $ 1/2 $ and $ 1 $ which leads to 
a phase doubling in half period of modulation. 

\begin{figure}[!ht]
\includegraphics[width=\textwidth,keepaspectratio]{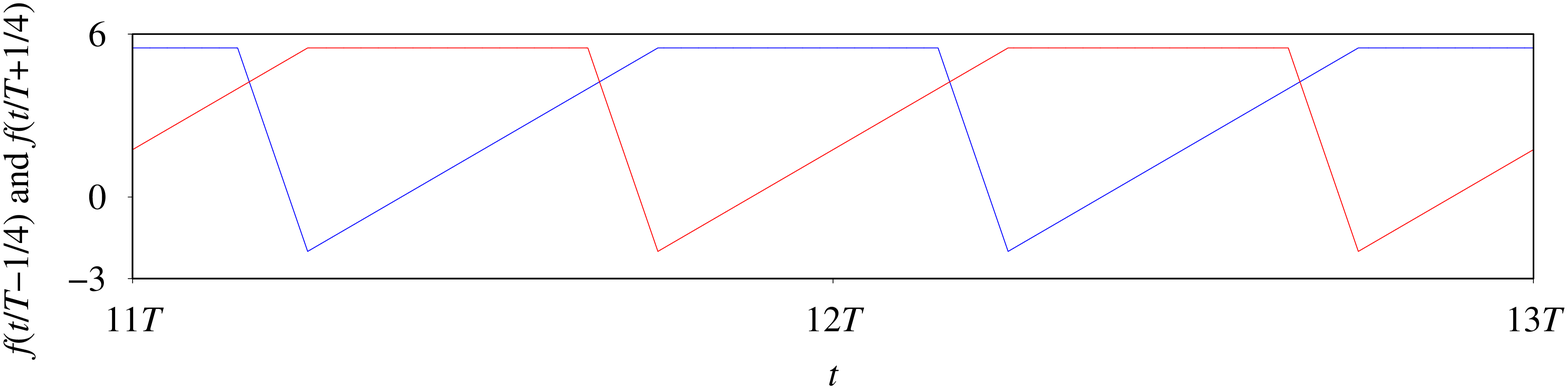}
\includegraphics[width=\textwidth,keepaspectratio]{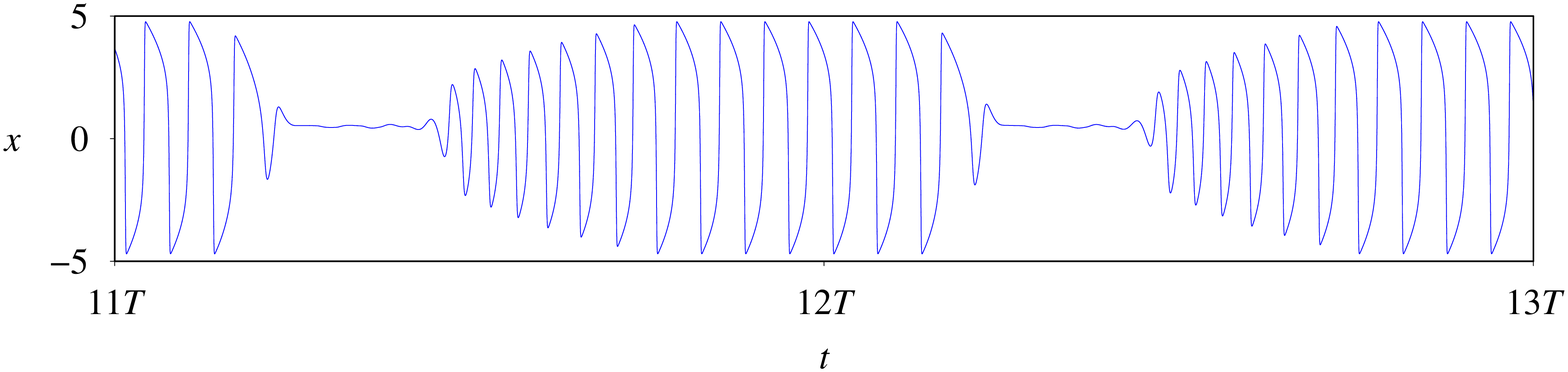}
\includegraphics[width=\textwidth,keepaspectratio]{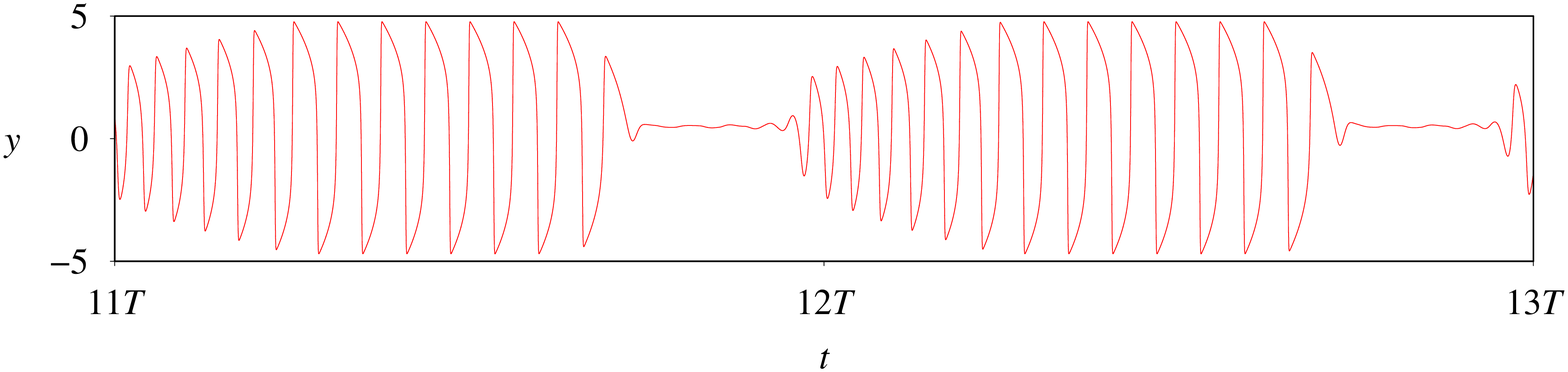}
\caption{The functions governing the modulation of the control parameters and waveforms of the variables \textit{x} and \textit{y}, 
$ a=5.49 $, $ K=0.5 $, $ c=-2 $, $ \varepsilon=0.01 $, $ T=200 $, $ \tau _1=0.4 $, $ \tau _2=0.5 $. }
\label{fig6}
\end{figure}

\begin{figure}[!h]
\includegraphics[width=\textwidth,keepaspectratio]{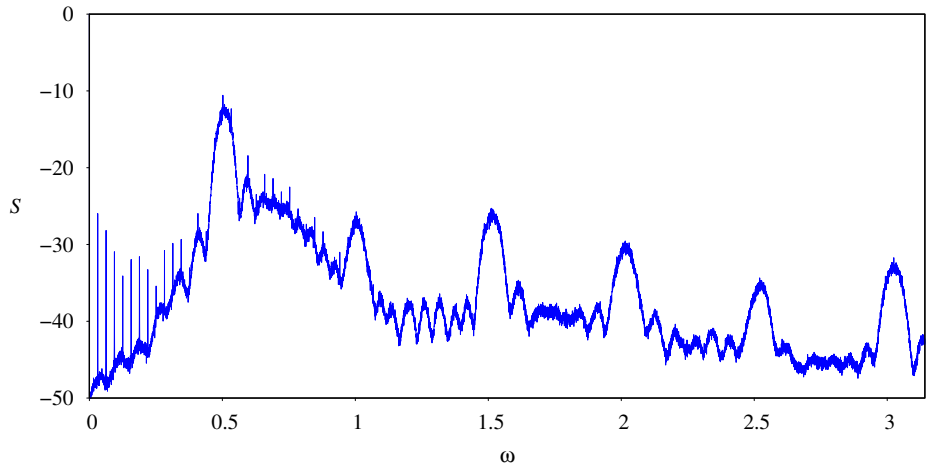}
\caption{The power density spectrum of the first oscillator in logarithmic scale at $ a = 5.49 $, $ K = 0.5 $, $ c = -2 $, 
$ \varepsilon = 0.01 $, $ T= 200 $, $ \tau _1 = 0.4 $, $ \tau _2 = 0.5 $.}
\label{fig7}
\end{figure}

Poincar\'e mapping over the modulation period was analyzed numerically to describe the dynamics of the phases in the system~\eqref{eq5} 
and to show presence of the Smale--Williams solenoid. Since the self-oscillations with high amplitudes differ significantly from 
sinusoidal, a calculation of the phase through the arctangent of the ratio of the variable and its derivative appears to be 
unsatisfactory. An alternative is to use a value that corresponds to a time shift from a given reference point normalized to a period of 
self-oscillations.

Let \textit{t} be the time of the start of excitation of the second oscillator, $ t_1 $ and $ t_2 $ are the preceding moments of the 
first oscillator passages through the section $ x = 0 $, where $ t_2 > t_1 $. Then it is possible to determine the angular (phase) 
variable belonging to the interval $ \left[ 0; \ 1 \right) $ as $ \varphi = \frac{t - t_2}{t_2 - t_1} $. Calculation of this quantity is 
easily programmed and performed during the numerical simulation of the dynamics of the system. 

\begin{figure}[!ht]
\includegraphics[width=.5\textwidth,keepaspectratio]{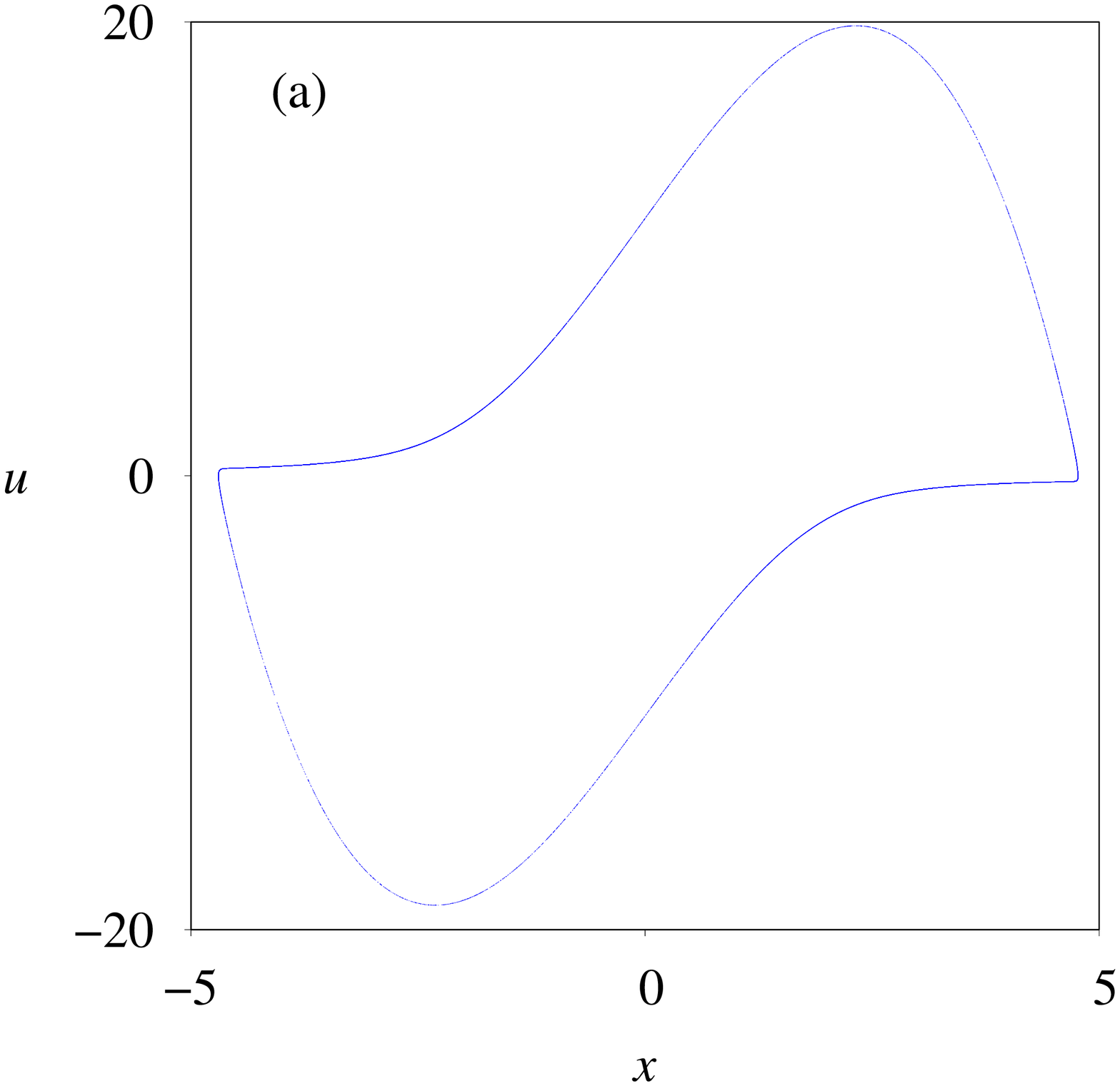}
\includegraphics[width=.5\textwidth,keepaspectratio]{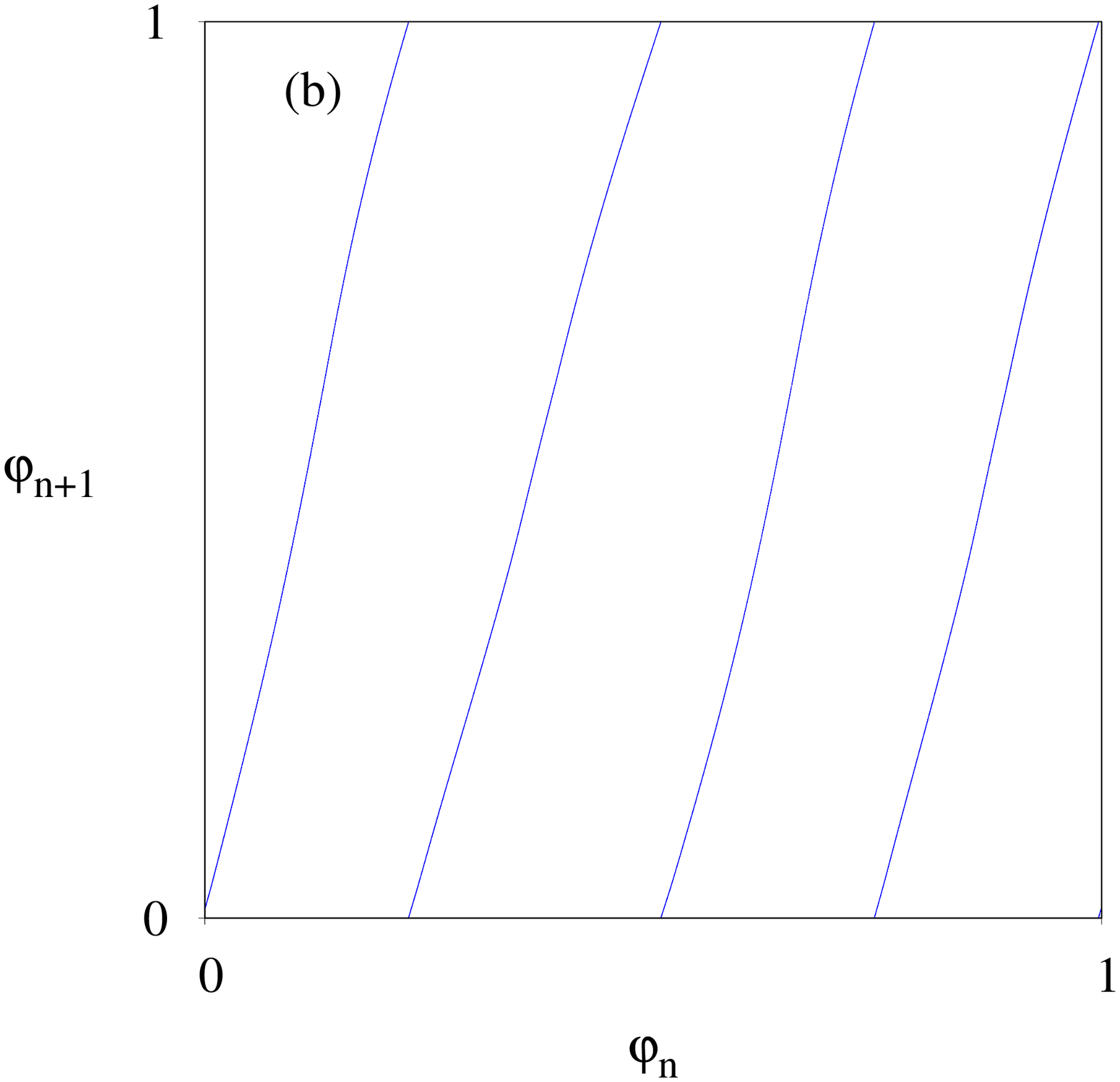}
\caption{Portrait of the Poincar\'e map attractor over the modulation period (in the projection onto the plane of the variables of 
the first oscillator) (a). Iteration diagram for the oscillation phase of the first oscillator (b). 
$ a = 5.49 $, $ K = 0.5 $, $ c = -2 $, $ \varepsilon = 0.01 $, $ T= 200 $, $ \tau _1 = 0.4 $, $ \tau _2 = 0.5 $. }
\label{fig8}
\end{figure}

Fig.~\ref{fig8} (a) and (b) depict a portrait of the Poincar\'e map attractor over the modulation period (in the projection onto the 
plane of the variables of the first oscillator) and the iteration diagram for the oscillation phase of the first oscillator at 
$ a = 5.49 $, $ K = 0.5 $, $ c = -2 $, $ \varepsilon = 0.01 $, $ T= 200 $, $ \tau _1 = 0.4 $, $ \tau _2 = 0.5 $. The iteration diagram 
for the phase corresponds to a Bernoulli map with factor $4$.

Lyapunov exponents were calculated by Benettin algorithm~\cite{17, 18} for the Poincar\'e map attractor over the modulation period. 
In accordance with the Benettin algorithm, the equations~\eqref{eq5} were integrated numerically simultaneously with 4 sets of equations 
for small perturbations:

\begin{equation}
\begin{aligned}
\delta \dot{x} &= \delta u, \\
\delta \dot{u} &= \left( f \left( \frac{t}{T} + \frac{1}{4} \right) -x^2 \right) \delta u - \left( 2 x u + 1 \right) \delta x + \varepsilon \left( \delta y - \delta x \right), \\ 
\delta \dot{y} &= \delta v, \\
\delta \dot{v} &= \left( f \left( \frac{t}{T} - \frac{1}{4} \right) -y^2 \right) \delta v - \left( 2 y v + 1 \right) \delta y + \varepsilon \left( \delta x - \delta y \right).
\label{eq7}
\end{aligned}
\end{equation}

Each set of the equations corresponds to a perturbation vector. These perturbation vectors were orthogonalized on each step of numerical 
solution with Gram--Schmidt algorithm. The logarithms of norms of vectors obtained in Gram-Schmidt procedure were accumulated in sums. 
These sums divided by the time of the simulation are estimations of the Lyapunov exponents. With the parameters 
$ a = 5.49 $, $ K = 0.5 $, $ c = -2 $, $ \varepsilon = 0.01 $, $ T= 200 $, $ \tau _1 = 0.4 $, $ \tau _2 = 0.5 $, the Lyapunov exponents 
(their mean values for 500 randomly selected trajectories of the attractor) are: $\Lambda _1 = 1.379 \pm 0.004 $, 
$\Lambda _2 = -34.45 \pm 0.07 $, $\Lambda _3 = -396.27 \pm 0.18 $, $\Lambda _4 = -1604.67 \pm 0.34 $.

The largest exponent is positive and is close to $ \ln 4 $ which is the Lyapunov exponent of the Bernoulli mapping with the stretching 
factor $4$. Other exponents are negative. This corresponds to the Smale--Williams solenoid with a four-time extension of the angular 
variable over one iteration of the Poincar\'e map.

\begin{figure}[!ht]
\includegraphics[width=\textwidth,keepaspectratio]{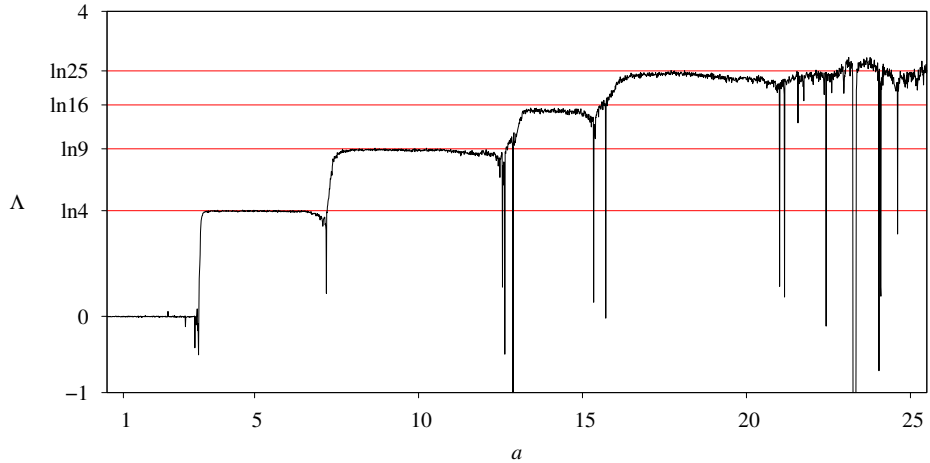}
\caption{The dependence of the highest Lyapunov exponent on the parameter $ a $ at $ K = 0.5 $, $ c = -2 $, 
$ \varepsilon = 0.01 $, $ T= 400 $, $ \tau _1 = 0.4 $, $ \tau _2 = 0.5 $.}
\label{fig9}
\end{figure}

Also, it is possible to implement Smale-Williams solenoids with other factors of the angular variable extension in the system~\eqref{eq5}. 
Fig.~\ref{fig9} shows the dependence of the highest Lyapunov exponent on the parameter a at $ T = 400 $, $ K = 0.5 $, $ c = -2 $, 
$ \varepsilon = 0.01 $, $ \tau _1 = 0.4 $, $ \tau _2 = 0.5 $. The highest Lyapunov exponent takes values corresponding to the expanding 
of the angular variable in $4$, $9$, $16$, $25$ times. As an example, Fig.~\ref{fig10} shows the iterative diagram of the Poincar\'e map 
over a half-period of modulation (this mapping is easily iterated numerically) at $ a = 17.5 $, $ T = 400 $, $ K = 0.5 $, $ c = -2 $, 
$ \varepsilon = 0.01 $, $ \tau _1 = 0.4 $, $ \tau _2 = 0.5 $. The diagram corresponds to the Bernoulli map with 5-fold expansion (for a 
full modulation period, we would have a 25-fold expansion diagram, which is not very convenient to visualize).

\begin{figure}[!ht]
\centering
\includegraphics[width=.5\textwidth,keepaspectratio]{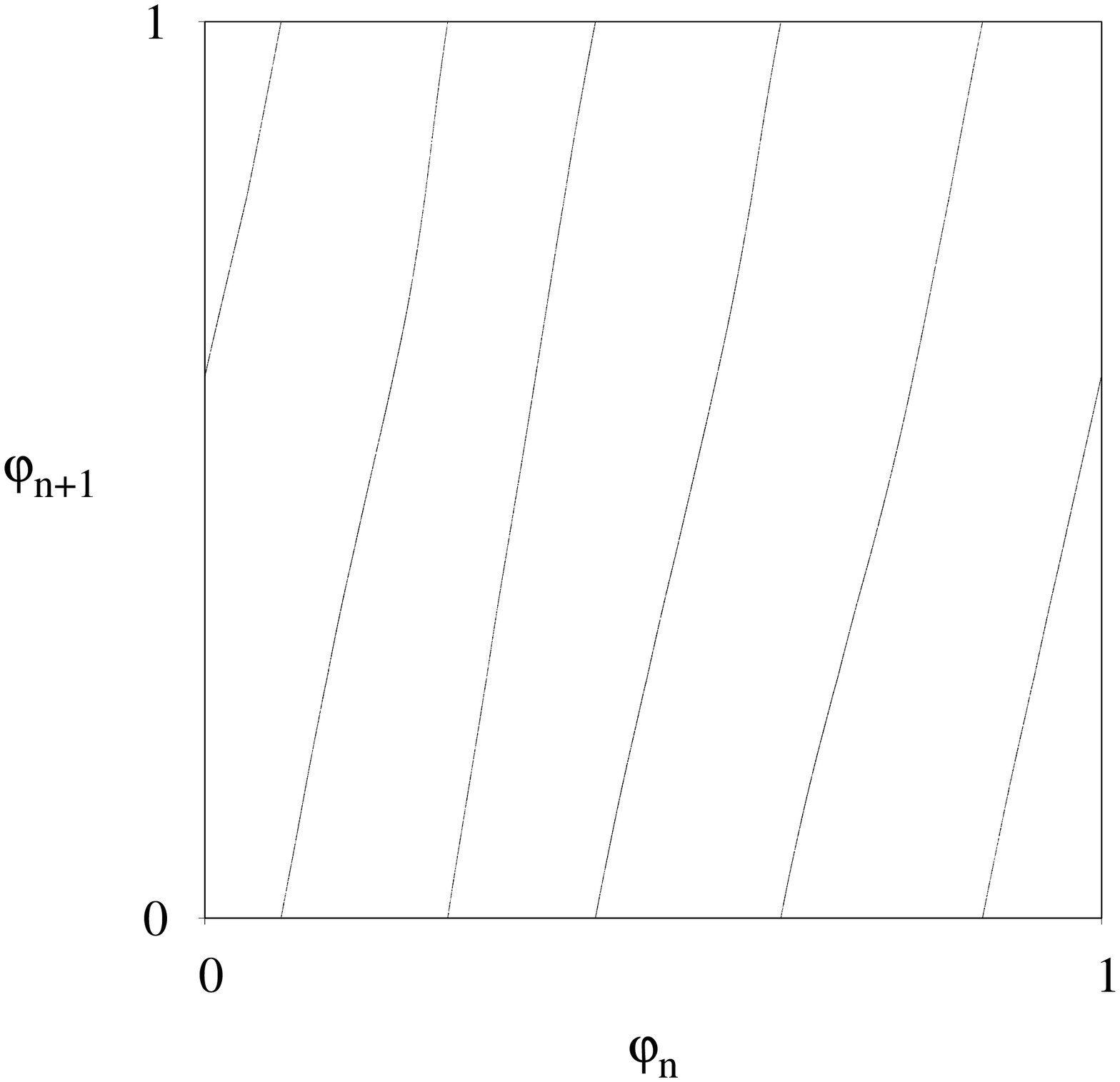}
\caption{The iterative diagram of the Poincaré map over a half-period of modulation at the values of the parameters 
$ a = 17.5 $, $ K = 0.5 $, $ c = -2 $, $ \varepsilon = 0.01 $, $ T= 400 $, $ \tau _1 = 0.4 $, $ \tau _2 = 0.5 $.}
\label{fig10}
\end{figure}

\begin{figure}[!ht]
\centering
\includegraphics[width=.8\textwidth,keepaspectratio]{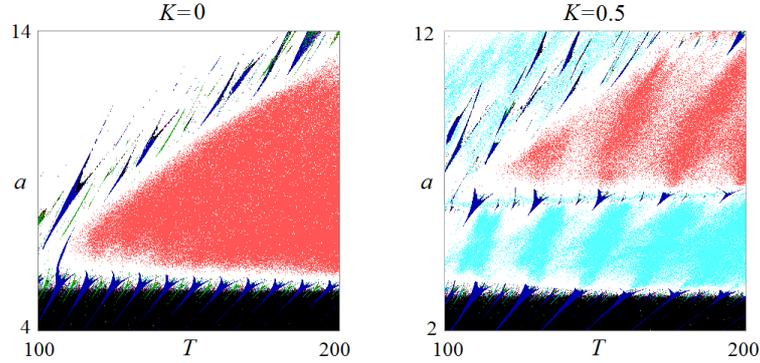}
\caption{Charts of regimes of the system~\eqref{eq5} on the parameter plane $ \left( T, \ a \right) $ for the values of the parameter 
$ K = 0 $ and $ K = 0.5 $; $ \varepsilon = 0.01 $, $ \tau _1 = 0.4 $, $ \tau _2 = 0.5 $. The regimes with the Smale--Williams solenoid 
with expanding factor $4$ are marked blue, Smale--Williams solenoids with factor $9$ are marked red, quasi-periodic regimes are black, 
non-hyperbolic chaotic regimes and periodic regimes are also distinguished. }
\label{fig11}
\end{figure}

\begin{figure}[!h]
\centering
\includegraphics[width=.4\textwidth,keepaspectratio]{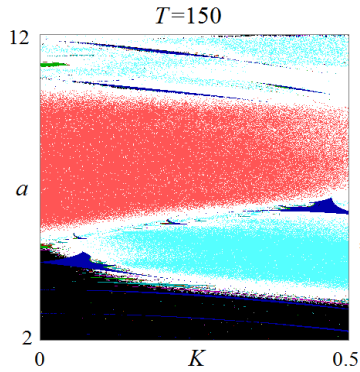}
\caption{A chart of regimes of the system~\eqref{eq5} in the parameter plane $ \left( K, \ a \right) $ at 
$ T = 150 $, $ c = -2 $, $ \varepsilon = 0.01 $, $ \tau _1 = 0.4 $, $ \tau _2 = 0.5 $. The color coding is similar to Fig.~\ref{fig11}. }
\label{fig12}
\end{figure}

Fig.~\ref{fig11} demonstrates charts of regimes of the system~\eqref{eq5} on the parameter plane $ \left( T, \ a \right) $ for the values 
of the parameter $ K = 0 $ and $ K = 0.5 $; $ c = -2 $, $ \varepsilon = 0.01 $, $ \tau _1 = 0.4 $, $ \tau _2 = 0.5 $. The regimes were 
defined by the value of the highest Lyapunov exponent. The regimes with the Smale--Williams solenoid with expanding factor $4$ are marked 
blue, Smale--Williams solenoids with factor $9$ are marked red, quasi-periodic regimes are black, non-hyperbolic chaotic regimes and 
periodic regimes are also distinguished. At $ K = 0 $ Smale--Williams solenoids with factor $4$ are absent. Fig.~\ref{fig12} shows charts 
of regimes of the system~\eqref{eq5} in the parameter plane $ \left( K, \ a \right) $ at $ T = 150 $, $ c = -2 $, $ \varepsilon = 0.01 $, 
$ \tau _1 = 0.4 $, $ \tau _2 = 0.5 $. At small $ K $ regimes with 4-fold phase expansion are not observed.

Numerical verification of hyperbolicity was performed, based on the transversality of stable and unstable manifolds of a hyperbolic 
attractor~\cite{19, 20, 21, 22}. In the case of Poincaré map of system~\eqref{eq5} an unstable manifold of typical trajectory of chaotic 
attractor is one-dimensional and one can compute vectors 
$ \delta \textbf{X} _n = \begin{bmatrix} \delta x_n, \ \delta u_n, \ \delta y_n, \ \delta v_n, \ \end{bmatrix} $ tangent to it by simultaneous solution of 
equations~\eqref{eq5} and variational equations~\eqref{eq7} in forward time (with normalization to avoid numerical register overflow). 
A stable manifold is three-dimensional but instead of computing three vectors that span it one can find the vector 
$ \delta \tilde{\textbf{X}} _n = \begin{bmatrix} \delta \tilde{x}_n, \ \delta \tilde{u}_n, \ \delta \tilde{y}_n, \ \delta \tilde{v}_n, \ \end{bmatrix} $ 
orthogonal to it by integrating the adjoint system of variational equations in the backward time along the same trajectory~\cite{21}. 
The adjoint system is a set of variational equations obtained by transposing Jacobi matrix and reversing the sign of all its elements:

\begin{equation}
\begin{aligned}
\delta \dot{\tilde{x}} &= \left( 2 x u + 1 \right) \delta \tilde{u} + \varepsilon \left( \delta \tilde{v} - \delta \tilde{u} \right), \\
\delta \dot{\tilde{u}} &= -\delta \tilde{x} - \left( f \left( \frac{t}{T} + \frac{1}{4} \right) -x^2 \right) \delta \tilde{u}, \\ 
\delta \dot{\tilde{y}} &= \left( 2 y v + 1 \right) \delta \tilde{v} + \varepsilon \left( \delta \tilde{u} - \delta \tilde{v} \right), \\
\delta \dot{\tilde{v}} &= -\delta \tilde{y} - \left( f \left( \frac{t}{T} - \frac{1}{4} \right) -y^2 \right) \delta \tilde{v}. 
\label{eq8}
\end{aligned}
\end{equation}

The scalar product of the vectors $ \delta \textbf{X} _n $ and $ \delta \tilde{\textbf{X}} _n $ defined by the equations~\eqref{eq7} 
and~\eqref{eq8} should be constant: 

\begin{equation}
\delta \textbf{X} _n \cdot \delta \tilde{\textbf{X}} _n = \delta x_n \delta \tilde{x} _n + \delta u_n \delta \tilde{u} _n + 
\delta y_n \delta \tilde{y} _n + \delta v_n \delta \tilde{v} _n = \text{const}.
\label{eq9}
\end{equation}

With this we obtained sets of vectors for sufficiently long trajectory of Poincaré map of system~\eqref{eq5}. At every point of the 
trajectory we calculated angles of intersections of the stable and unstable manifolds 
$ \alpha _n = \frac{\pi}{2} - \arccos \frac{\delta \textbf{X} _n \cdot \delta \tilde{\textbf{X}} _n}{\| \delta \textbf{X} _n \| \ \| \delta \tilde{\textbf{X}} _n \| } $. 
Figs.~\ref{fig13}--\ref{fig16} show histograms of the distributions of the angles between stable and unstable manifolds for 
trajectory of attractor of the Poincar\'e map of system~\eqref{eq5} in four regimes.

The histograms on Figs.~\ref{fig13}--\ref{fig15} correspond to regimes with hyperbolic attractors with expanding factors 
$ M=4 $ and $25$ $ (K = 0.5) $ and $9$ $ (K = 0) $. In all three regimes zero angles are absent and the minimum angle is distanced from 
zero. In contrast, the histogram on Fig.~\ref{fig16} corresponds to a nonhyperbolic chaotic attractor manifesting occurrence of the 
angles close to zero with notable probability.

\begin{figure}[!ht]
\centering
\includegraphics[width=\textwidth,keepaspectratio]{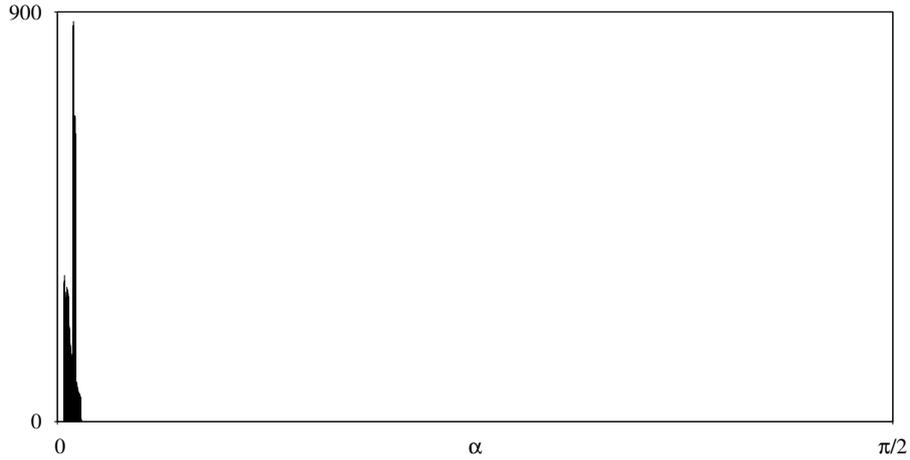}
\caption{Histogram of distribution of angles between manifolds at $ a = 5.49 $, $ K = 0.5 $, $ c = -2 $, $ \varepsilon = 0.01 $, 
$ T = 200 $, $ \tau _1 = 0.4 $, $ \tau _2 = 0.5 $. }
\label{fig13}
\end{figure}

\begin{figure}[!h]
\centering
\includegraphics[width=\textwidth,keepaspectratio]{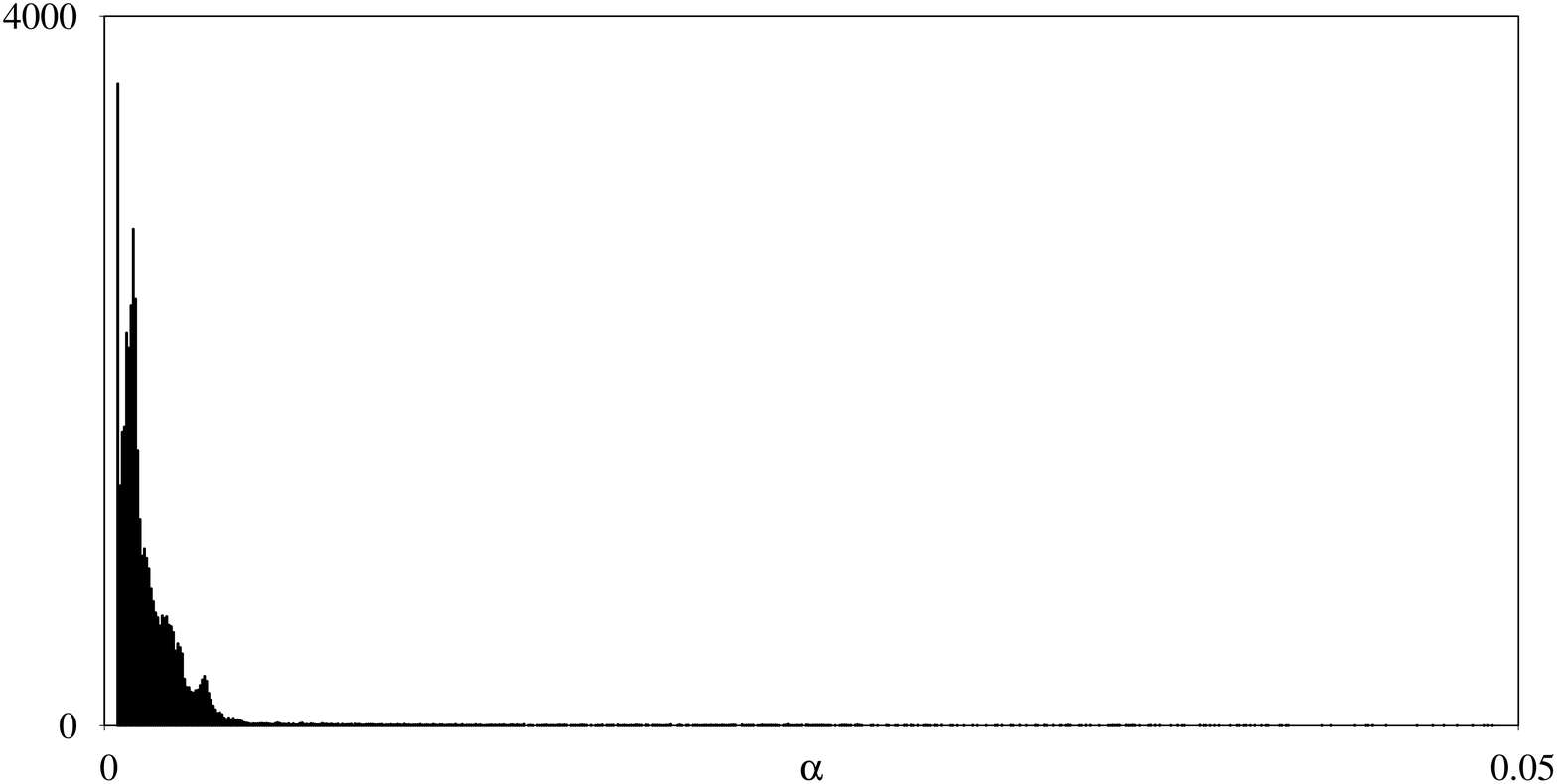}
\caption{Histogram of distribution of angles between manifolds at $ a = 17.5 $, $ K = 0.5 $, $ c = -2 $, $ \varepsilon = 0.01 $, 
$ T = 200 $, $ \tau _1 = 0.4 $, $ \tau _2 = 0.5 $. }
\label{fig14}
\end{figure}

\begin{figure}[!h]
\centering
\includegraphics[width=\textwidth,keepaspectratio]{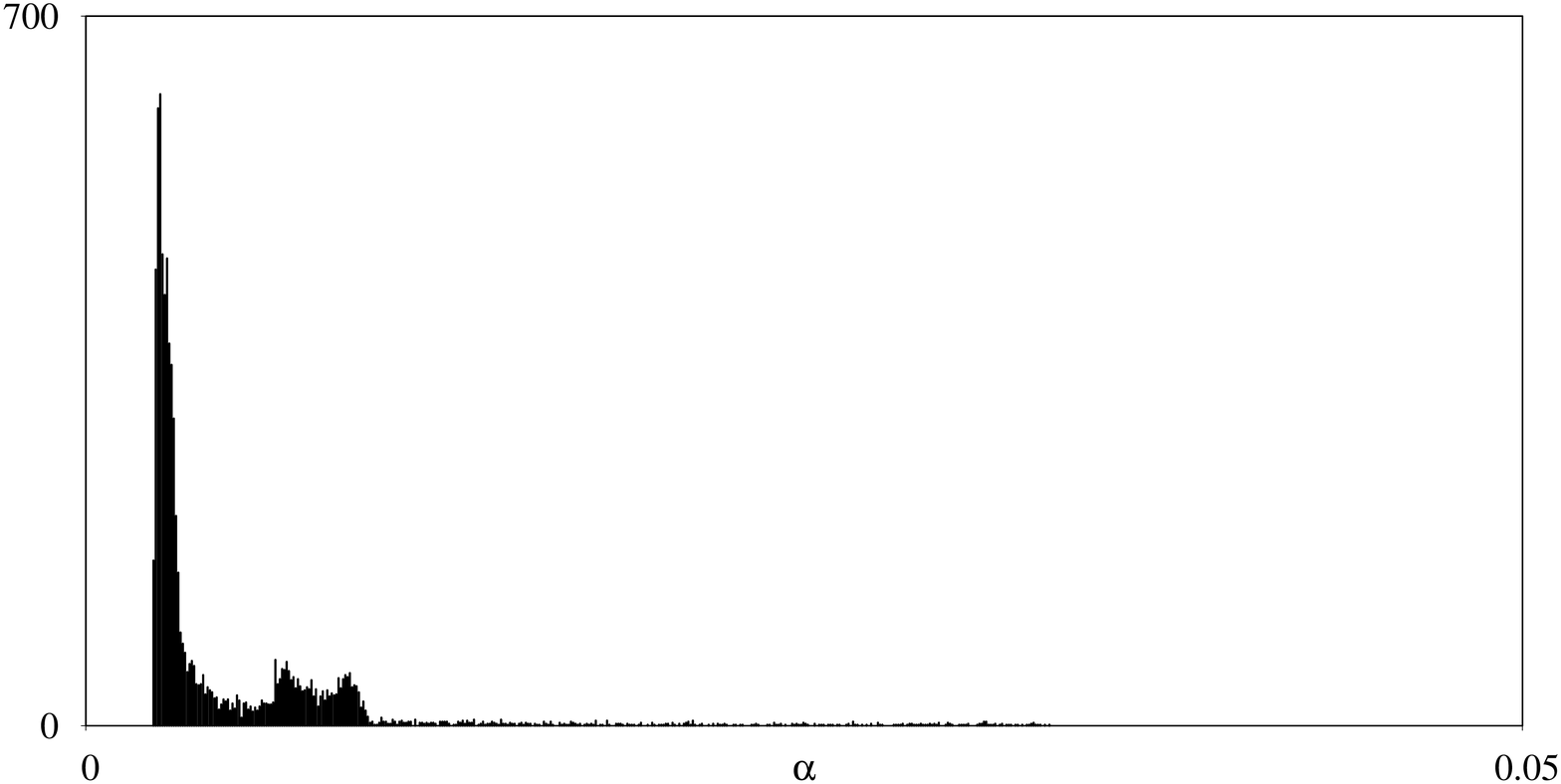}
\caption{Histogram of distribution of angles between manifolds at $ a = 5.49 $, $ K = 0 $, $ c = -2 $, $ \varepsilon = 0.01 $, 
$ T = 200 $, $ \tau _1 = 0.4 $, $ \tau _2 = 0.5 $. }
\label{fig15}
\end{figure}

\begin{figure}[!h]
\centering
\includegraphics[width=\textwidth,keepaspectratio]{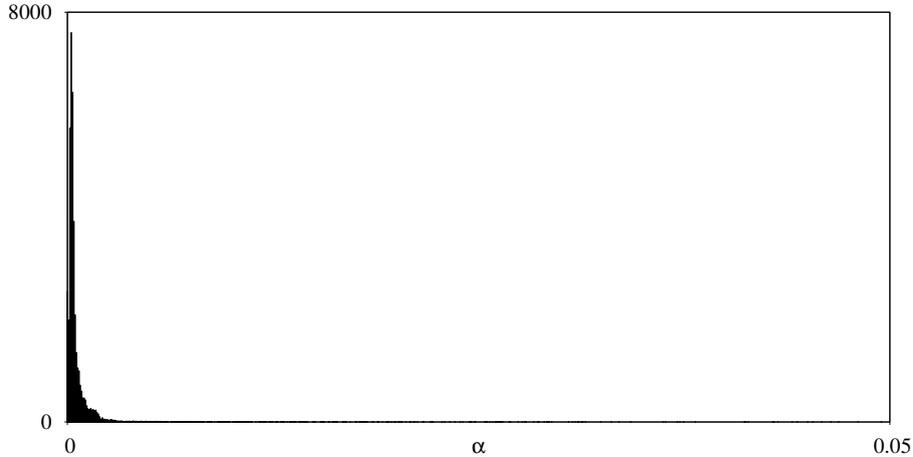}
\caption{Histogram of distribution of angles between manifolds at $ a = 25 $, $ K = 0.5 $, $ c = -2 $, $ \varepsilon = 0.01 $, 
$ T = 200 $, $ \tau _1 = 0.4 $, $ \tau _2 = 0.5 $. }
\label{fig16}
\end{figure}

\section{Circuit Implementation}

Let us turn to a circuit implementation of hyperbolic chaotic attractor in a system very similar to~\eqref{eq5}. The circuit consists of 
two symmetric alternately excited self-oscillating elements. There is a resonance transfer of excitation between them with phase 
doubling due to exploiting of the second harmonic of relaxation oscillations. 

Fig.~\ref{fig17} shows a circuit composed of two identical self-oscillating elements based on the contours $ \text{L}_1 \text{C}_1 $ 
and $ \text{L}_2 \text{C}_2 $. The negative resistance in one and the other contour is provided by the operational amplifiers 
$ \text{OA}_1 $ and $ \text{OA}_2 $. A value of the negative resistance depends on an instantaneous drain-source resistance of the 
field effect transistors $ \text{Q}_1 $ and $ \text{Q}_2 $. The control voltage supplied from the $ \text{V}_1 $ and $ \text{V}_2 $ 
sources remains zero during a certain part of the modulation period (the oscillator is active). For the rest of the period the voltage 
is less than zero, its time dependence has the form of a triangular function, and the oscillations are suppressed. The control voltages 
of the first and second subsystems are shifted to each other in time by a half-period of modulation. The parameters of the negative 
resistance elements are chosen so that the zero gate voltage corresponds to arising relaxation oscillations with half the frequency of 
the linear oscillations. The diodes ($ \text{D}_1 $ and $ \text{D}_2 $) are included in each oscillation circuit; it provides a 
saturation of the amplitude and a generation of intense second harmonic of large amplitude oscillations. When the next stage of the 
oscillator's activity comes, the growth of the oscillations, starting from small amplitude, is effectively stimulated resonantly by the 
second harmonic of the oscillations of the partner oscillator due to the coupling through the capacitor $ \text{C}_3 $, while the partner 
oscillator is just at the stage of high amplitude relaxation oscillations. Next, the second oscillator becomes suppressed, whereas the 
first demonstrates relaxation oscillations, and then the process repeats with alternating exchange of the roles of both oscillators. As 
the excitation is transmitted from one oscillator to the other via the second harmonic, it goes with the phase doubling, which with 
contraction along the remaining directions in the state space ensures formation of the attractor of Smale-Williams type in the 
stroboscopic mapping over the period of modulation. 

\begin{figure}[!h]
\centering
\includegraphics[scale=.5,keepaspectratio]{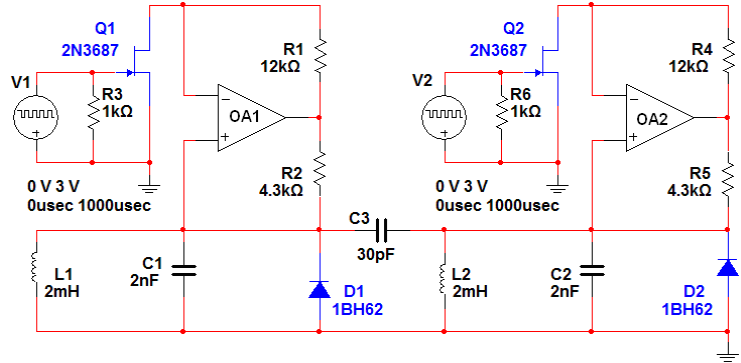}
\caption{A system of two coupled self-oscillating elements realizing hyperbolic chaos. }
\label{fig17}
\end{figure}

\begin{figure}[!h]
\centering
\includegraphics[width=\textwidth,keepaspectratio]{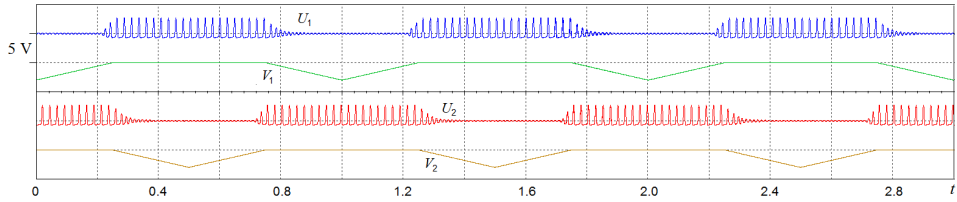}
\caption{Waveforms of voltages $ \text{U}_1 $ and $ \text{U}_2 $ on capacitors $ \text{C}_1 $ and $ \text{C}_2 $ and control voltages 
on the gate of field effect transistors $ \text{V}_1 $ and $ \text{V}_2 $, obtained with the help of a virtual four-beam oscilloscope in 
the Multisim simulation. The nominal values of the circuit elements correspond to Fig.~\ref{fig17}. }
\label{fig18}
\end{figure}

Fig.~\ref{fig18} illustrates operation of the circuit in accordance with the described mechanism. The oscilloscope traces were obtained 
in the Multisim simulation using the virtual four-beam oscilloscope.
 
Confirmation of the hyperbolic nature of chaos requires us to verify that the successive stages of activity correspond to the phase 
transformation according to the Bernoulli map. Phase may be defined as a time shift with respect to a given reference point, normalized 
to the characteristic period of the relaxation oscillations (as in the previous sections). To depict graphically the quantities related 
to successive stages of the activity one can use the simulation data recorded in file from Multisim at a sufficiently long time with a 
small sampling step (less than the period of small oscillations) using the Grapher tool. A diagram obtained by processing such data is 
shown in Fig.~\ref{fig19} (a). Since the doubling of the phase variable occurs with each transmission of the excitation from one 
oscillator to another, the phase expands with factor $4$ during each modulation period, and the plot consists of 4 branches inclined 
with the slope coefficient close to 4. Diagrams (b), (c) of Fig.~\ref{fig19} show the spectrum of the signal generated by one of the 
oscillators in linear and logarithmic scales. The spectrum is jagged due to the fact that the signal has the form of periodically running 
oscillation trains; however the spectrum is certainly continuous because of the chaotic nature of the phase shift at successive stages of 
activity. 

\begin{figure}[!ht]
\includegraphics[width=.33\textwidth,keepaspectratio]{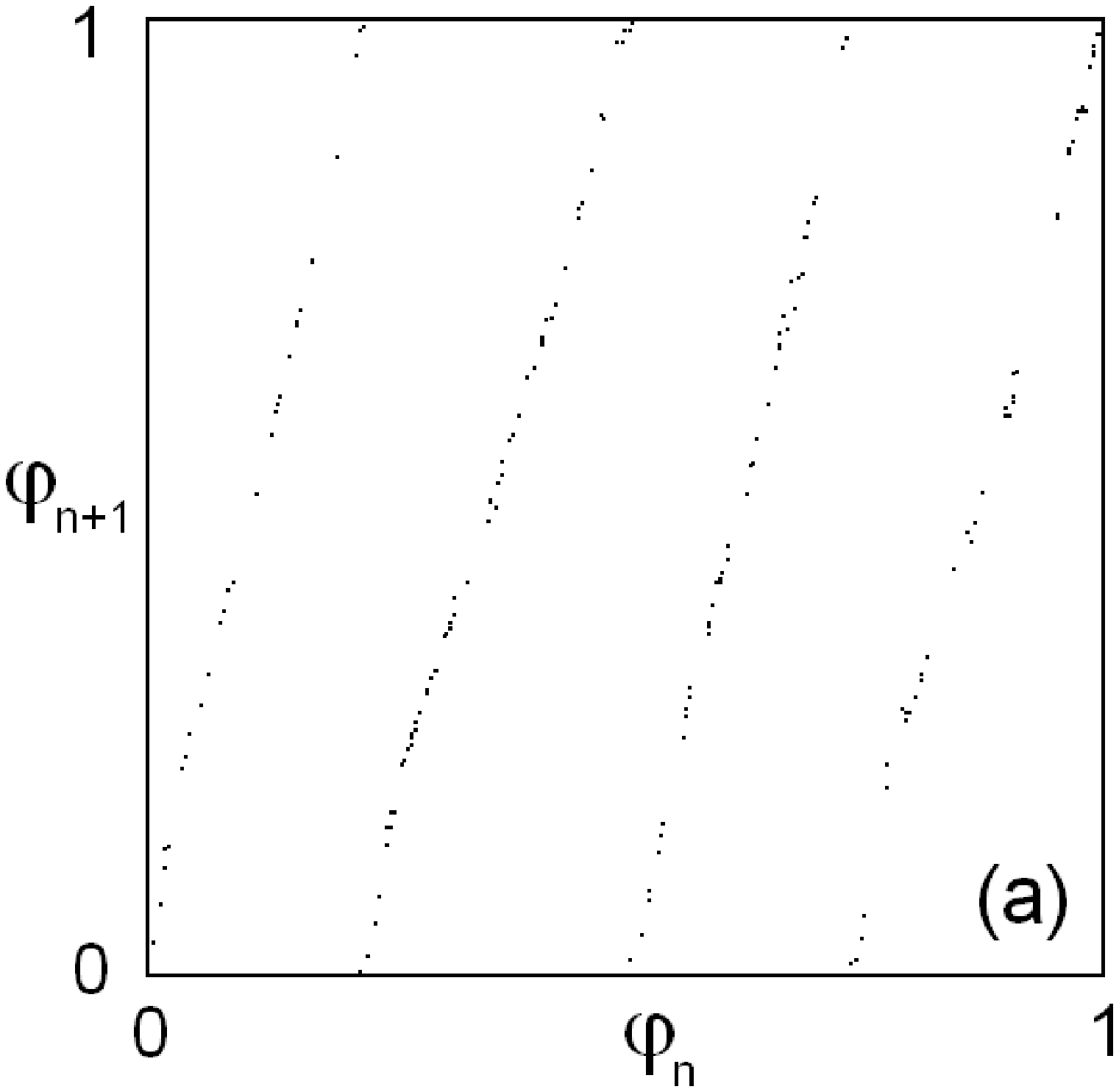}
\includegraphics[width=.67\textwidth,keepaspectratio]{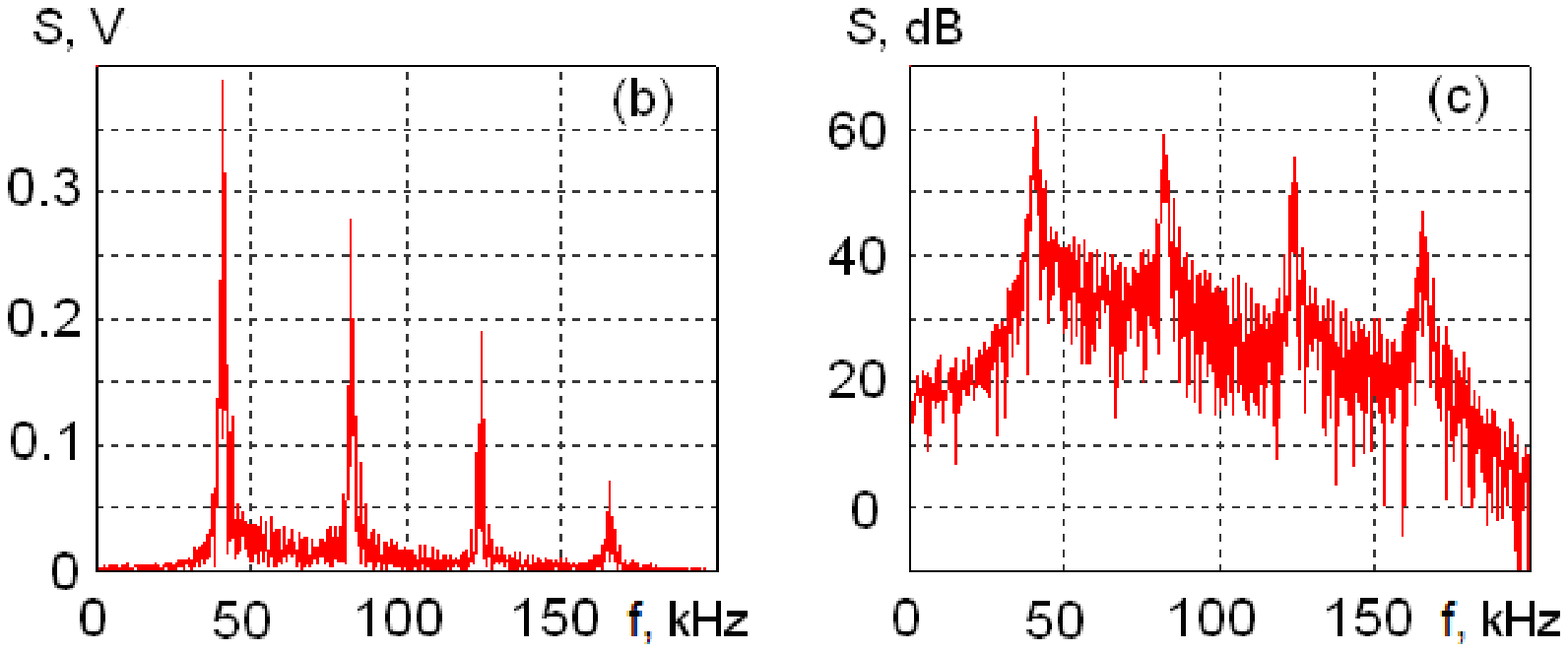}
\caption{A diagram illustrating the phase transformation in successive stages of activity (a) and the oscillation spectra in linear (b) 
and logarithmic scale (c) for one of the oscillators as obtained in the Multisim simulation of the circuit in Fig.~\ref{fig17}. }
\label{fig19}
\end{figure}

\section{Conclusion}

We discuss a new approach in constructing systems with hyperbolic attractors. The idea is to modulate the control parameters of coupled 
self-oscillators in such way that their frequencies vary in integer number of times. With parameters selected correctly the excitation 
transfers resonantly between the oscillators. The phase of excitation undergoes the Bernoulli map on full revolution of excitation.

We have proposed an example of system based on new principle. The system is composed of two weakly coupled Bonhoeffer--van der Pol 
oscillators with periodic modulation of control parameters. A mathematical model is derived and its numerical simulation is performed. 
It is demonstrated that Smale--Williams solenoids with different expanding factors appear in Poincar\'e stroboscopic map on specific 
intervals of parameter values. The hyperbolicity of chaotic attractors is confirmed with the help of a test based on numerical 
evaluation of angles of intersections of stable and unstable manifolds of attractor with verification of the absence of tangencies 
between these manifolds. The system can be implemented as electronic circuit. We have studied a scheme based on our mathematical model 
in Multisim simulation. It demonstrated the dynamics on Smale--Williams attractor.

\textit{The work was supported by grant of RSF No 17-12-01008 (sections 1-3, formulation of the model, numerical simulation and analysis) 
and by grant of RFBR No 16-02-00135 (section 4, circuit implementation and Multisim simulation).}

}
\label{article_end}

\end{document}